\documentclass{article}
\usepackage{amsmath}
\usepackage{amscd}
\usepackage{latexsym}
\usepackage{amssymb}

\usepackage{color}

\title{Path algebras and de Broglie waves}
\author{Murray Gerstenhaber}

\begin{document}
\maketitle
\newtheorem{theorem}{Theorem}
\newtheorem{corollary}{Corollary}
\newtheorem{lemma}{Lemma}
\renewcommand{\abstractname}{}
\newcommand{\C}{\ensuremath{\mathbb{C}}}
\newcommand{\R}{\ensuremath{\mathbb{R}}}
\newcommand{\Z}{\ensuremath{\mathbb{Z}}}
\newcommand{\cP}{\ensuremath{\mathcal{P}}}
\newcommand{\cM}{\ensuremath{\mathcal{M}}}
\newcommand{\I}{\ensuremath{\mathcal{I}}}
\newcommand{\bM}{\ensuremath{\mathbf{M}}}
\newcommand{\cX}{\ensuremath{\mathcal{X}}}
\newcommand{\cA}{\ensuremath{\mathcal{A}}}
\newcommand{\bA}{\ensuremath{\mathbf{A}}}
\newcommand{\cB}{\ensuremath{\mathcal{B}}}
\newcommand{\cS}{\ensuremath{\mathcal{S}}}
\newcommand{\pr}{\ensuremath{\preceq}}
\newcommand{\op}{\ensuremath{\mathrm{op}}}
\newcommand{\bfk}{\ensuremath{\mathbf{k}}}
\newcommand{\g}{\gamma}
\newcommand{\G}{\Gamma}
\newcommand{\z}{\zeta}
\newcommand{\Aut}{\operatorname{Aut}}
\newcommand{\dr}{\operatorname{dR}}
\vspace{-7mm}
\date{}
{}

\begin{abstract}\noindent 
De Broglie waves may be a reflection of a deformation inherent in the path algebra of phase space. On a Riemannian manifold equipped with a suitable closed 2-form, the product of paths, which is ordinarily their concatenation, can be deformed by multiplication by a scalar weight giving rise to a function on paths. In flat phase space the associated function is periodic with period the de Broglie wave length. The de Broglie description may only be approximate in curved space.


\end{abstract}

\section{Introduction}
Wave-particle duality assigns to every moving body its de Broglie wave length $h/p$, where $h$ is Planck's constant and $p$ its momentum. This paper applies algebraic deformation theory to the multiplication of paths on a Riemannian manifold carrying a suitable closed 2-form, and in particular to phase space. The ordinary product of paths is their concatenation, but this may be deformed in a way that defines a weight or phase function on the path.   The de Broglie wave of a particle may be a reflection of such a deformed product of paths in phase space. If so, the de Broglie formula, exact in a flat space, may be different when the space is curved, or equivalently, when a particle is acted upon by a force,  cf \S \ref{sphere}.

In the  simplest case suppose that one has a flat space with an area element  $\omega$. The ordinary product of a straight line segment $[ab]$ with a second  $[bc]$ would be the broken line segment $[abc]$. If $\alpha$ is the area of the triangle spanned by $a,b,c$ (in that order) then the deformed product is of the form $\exp(2\pi i\alpha/\tau)[abc]$ where $\tau$ is a certain ``taille''. The weight function here is $\exp(2\pi i\alpha/\tau)$. Suppose now that a particle moves along the ray starting at $b$ and in the direction of $c$ and that its position at time $t$ is $c(t)$ with $c(0) = b$. Let the area of the triangle spanned by $a, b, c(t)$ be $\alpha(t)$.  The periodic function $\exp(2\pi i\alpha(t)/\tau)$ may be viewed as defining a wave on the ray. In the simplest phase space with  $[ab]$ in the momentum direction,  $[bc]$ in the position direction, take $\tau = h/2$.  If the length of $[ab]$ is $p$, corresponding to a particle instantaneously accelerated from rest to a momentum of p, then the wave length is the de Broglie wave length $h/p$.

Let $\omega$ denote a closed 2-form on a manifold $\cM$.  Any such form determines a morphism of the second homotopy group $\pi_2 = \pi_2(\cM)$ of $\cM$ into the real numbers $\R$; the morphism depends only on the cohomology class $\bar\omega$ of $\omega$.  If this image is  a discrete subgroup of $\R$ (which may be reduced to zero) then it must be of the form  $\tau\R$ for some $ \tau \ge 0$. The value of $\tau$ will then be called the \emph{taille}  of $\omega$ (and of  $\bar\omega$) and $\omega$ will be called a \emph{tailor form}. 

Some of what follows does not depend on having a metric on $\cM$, but suppose henceforth for convenience that it is Riemannian. This gives a functional on paths, namely length. Others may do, such as action. 

The product of paths, defined more precisely later, is loosely their concatenation, or zero if they can not be concatenated. The algebra defined by these products is the path algebra of $\cM$, denoted  $\cA$. Using $\omega$, the multiplication of paths on $\cM$ can now be deformed by the introduction of a weight function on the product. A deformation induced by a 2-form will multiply the ordinary concatenation product by a non-zero real or complex weight.  This weight will be one if the two paths are geodesics and either their concatenation remains a geodesic, the second just being a continuation of the first, or if the second just reverses the direction of the first (a reflection).

If $\tau(\omega) = 0$, then $\omega$ can induce a continuous family of deformations of $\cA$, while if $\tau(\omega) > 0$ then there is only a discrete family quantized by the integers. In the quantized case, e.g., phase space,  one is forced to take take complex exponentials and the weights  will have absolute value one; in the non-quantized case they can take on arbitrary values. 

 While the deformation of the path algebra $\cA(\cM)$ of $\cM$ induced by a tailor  form $\omega$ depends, up to algebraic isomorphism, only on the cohomology class  $\bar\omega$, the properties of the actual deformations induced by cohomologous forms may differ. The area form on the plane is exact and the deformation it induces therefore technically trivial. On a compact orientable Riemannian manifold, each cohomology class of forms contains a unique harmonic one, which is a natural choice of representative.  This  does not single out the area form on the plane, but it does so on its quotient by a lattice, the flat 2-torus, although they are locally isomorphic. 
 
The area form, which produces the de Broglie wave, is dual to the infinitesimal of the deformation of $\R[x,y]$ to the first Weyl algebra, which is essentially the algebra of observables on the simplest phase space. This raises the question, amongst others, of when the dual of a harmonic 2-form can be extended to an algebra deformation by a simple universal deformation formula, as is the case for the Weyl algebra.

Suppose that a particle or object has traveled a path $\g$ from $a$ to $b$ and now is traveling on a second path $\g'$ beginning at $b$, which we may suppose has been parameterized by its length from $b$. When the particle has covered a distance $x$ on $\g'$, arriving at a point $\g'(x)$, the weight that will be attached to the product of $\g$ and the segment of $\g'$ from $b$ to $\g'(x)$ will be a function of $x$ which depends also on the path $\g$ that the particle initially traversed. The path of a particle may be linear in configuration space but in phase space it will appear deflected  if there is a change in its momentum. In particular, if it starts from rest and is instantly accelerated, then the projection onto configuration space of the first part of its path in phase space will just be a point but in phase space one does have a concatenation of two paths.  A particle moving along a path which is not a geodesic will have a weight assigned to it as  it moves. This weight, which is 1 at the start of the path, is a function of that part of the path already traveled. The weights generally depend on the metric, but when the path is closed the weight assigned to the total path will depend only on the 2-form inducing the deformation.  For closed paths one recovers Bohr's hypothesis, \S \ref{Bohr}.

The concept of deformation is essential to understanding quantization, something first made explicit in the foundational paper \cite{BFFLS} introducing deformation quantization. (For substantial later developments and references cf. Sternheimer \cite{Sternheimer:20YearsAfter}.) 
Phase space carries a canonical closed 2-form. Its taille is zero if it is  viewed simply as a vector bundle, namely, the cotangent bundle of configuration space. However, the existence of Planck's constant $h$ indicates that it behaves as though it actually had a non-zero taille, namely $h$ or  some submultiple of it.  Physical phase space is more than just a vector bundle, in particular, its coordinates, position and momentum, do not commute. It is this lack of commutativity which forces  one to compute area integrals modulo the uncertainty that it introduces, the taille.  In \cite{BFFLS} it was shown that the non-commutativity could be viewed as  a reflection of a deformation of the commutative polynomial algebra generated by the coordinates of phase space. Using ideas introduced in \cite{G:Path}, here we present a dual view: The existence of de Broglie waves can be viewed as reflecting a deformation of its path algebra. 

\section{Path algebras} By a \emph{path} on the Riemannian manifold $\cM$ we will mean a piecewise regular map of a directed segment of the real line into $\cM$.  The image then has a  well-defined length $\ell$, so parameterizing the image by its length from the starting point gives a map $\g: [0,\ell] \to \cM, \ell \ge 0$ such that for all $0 \le x \le \ell$ the length of $\g([0,x])$ is exactly $x$. One can have $\ell = 0$, in which case the path is reduced to a point. Suppose that $\g: [0,\ell] \to \cM$ and $\g': [0,\ell'] \to \cM$ are two paths on $\cM$.  If $\g(\ell) = \g'(0)$ then we define their concatenation product $\g\g'$ by setting $\g\g'(x)= \g(x)$ for $0\le x \le \ell$ and $\g\g'(x) = \g'(x-\ell)$ for $\ell \le x \le \ell+\ell'$. The product is defined to be zero otherwise. 

With this multiplication, the paths on $\cM$ together with 0 form a \emph{semigroup} $\bM$, that is, a set with a single associative multiplication but which does not necessarily contain a unit element. (A semigroup with a unit element is called a monoid.)  If we have a semigroup $\bM$ and a commutative unital ring $\bfk$ then we can form the \emph{semigroup algebra} $\bfk\bM$ whose underlying module is the free $\bfk$-module generated by the elements of $\bM$ (the set of all formal finite sums of elements of $\bM$ with coefficients in $\bfk$), with multiplication defined by that in $\bM$. When $\bM$ is the semigroup of paths on $\cM$ and $\bfk = \R$ this is the \emph{path algebra} of $\cM$.  Those paths which are piecewise geodesic segments form a subalgebra. The  path algebra of $\cM$, here denoted $\cA = \cA(\cM)$, does not have a unit element. (It would have to be the sum of all the points of $\cM$, which as elements of the path algebra are orthogonal idempotents, but here we are allowing only finite sums. Classical path algebras, which have been studied in connection with graphs and quivers, generally do have units.) An algebra $A$ (over a field) having a multiplicative basis, i.e., a basis $\cB$ such that the product of any two elements of $\cB$ is again in $\cB$ or zero, is in particular a semigroup algebra. Matrix algebras, poset algebras, group algebras, and algebras of finite representation type over algebraically closed fields  are examples. (For the latter cf \cite{FiniteRep}.) 


The path algebra carries a natural topology when a certain restriction is understood.  For convenience, a path $\g$ on $\cM$ can also be viewed as a map $[0,\infty)\to \cM$ which is constant for all $x \ge \ell(\g)$ where $\ell(\g)$ is the length of $\g$. We could therefore topologize the space of paths with the usual compact-open topology except that we want also that paths which are close in the topology to have lengths that are also close. To this end, view a path as defining a map of $[0,\infty) \to \cM \times [0,\infty)$ in which the second component is the map sending a path to its length, and then take the compact-open topology.  The multiplication is not strictly continuous with this topology since a product $\g\g'$ may by non-zero but the products of paths converging to $\g$ and $\g'$, respectively will always be zero if they can not be concatenated.  However, the multiplication is continuous when restricted to pairs of paths whose products are not zero, and these are the only products that are of interest. In this restricted sense $\cA(\cM)$ is a topological algebra. 

\section{Semigroup cohomology} What we will call the \emph{absolute} cohomology of a semigroup $\bM$ with coefficients in an additive group $\G$ is defined as follows. Let $\bM^n = \bM \times  \cdots\times \bM$ ($n$ times).  The $n$-cochains of $\bM$ with coefficients in $\G$ are mappings $F\!:\! \bM^n \to \G$; these form an additive group $C^n(\bM, \G)$. The coboundary operator $\delta: C^n \to C^{n+1}$ is defined by setting 
\begin{multline*}
\delta F(a_1,\dots,a_{n+1}) = F(a_2,\dots,a_{n+1})\, +\\ \sum_{i=1}^{n}(-1)^i F(a_1,\dots,a_{i-1}, a_ia_{i+1},a_{i+2},\dots,a_{n+1}) \\+ (-1)^{n+1}F(a_1,\dots,a_n).
\end{multline*}
Then $\delta\delta = 0$, the group $Z^n$ of $n$-cocycles is the kernel of $\delta$ on $C^n$, the subgroup of $n$-coboundaries $B^n$ is $\delta C^{n-1}$, and the $n$th cohomology group is $H^n(\bM, \G) = Z^n/B^n$. There are no 0-cochains, and unlike group cohomology, there is no operation of $\bM$ on the coefficient group $\G$. The cup sum of an $m$-cochain $F^m$
with an $n$-cochain $G^n$ is defined  by setting $(F^m\uplus G^n)(a_1,\dots, a_{m+n}) = F^m(a_1,\dots,a_m) + G^n(a_{m+1},\dots,a_{m+n})$.
One has
$$\delta(F^m \uplus G^n) = (\delta F^m) \uplus G^n + (-1)^mF^m \uplus \delta G^n,$$
so the cup sum of cocycles is a cocycle, that of a cocycle and a coboundary is a coboundary, and the cup sum is defined on the cohomology.   

A zero element in a semigroup $\bM$, generally denoted simply by 0, is an element such that $a\cdot 0 = 0\cdot a = 0$ for  $a \in \bM$.   When, as in the path semigroup, there is a zero element we will always restrict attention to the subcomplex consisting of those cochains $F\in C^n(\bM, \G)$ such that $F(a_1,a_2, \dots, a_n) = 0$ whenever $a_1a_2\cdots a_n = 0$ and call its cohomology simply the cohomology of the semigroup. This is necessary since, for example, without the restriction a 1-cocycle would simply be a function such that $F(ab) =F(a) +F(b)$, implying that $F(a) = -F(b)$ whenever $ab = 0$.  For the path algebra it would follow that $F$ must be identically zero, whereas using the subcomplex the requirement is that $F(ab) = F(a) + F(b)$ when $a$ and $b$ can be concatenated and is zero otherwise, which is what is really wanted. This restricted cohomology has been called the ``0-cohomology'' by Novikov, cf \cite{Novikov:Semigroup}.

When the coefficient group $\G$ is multiplicative the coboundary formula can be rewritten in multiplicative form and the cup sum becomes the usual cup product.  The condition that a multiplicative 1-cochain $f$ be a cocycle is then that $f(ab) = f(a)f(b)$ when $ab \ne 0$.  (The coefficients now actually need  only form a commutative semigroup.)  A 2-cocycle is a cochain $f$ such that   $f(a,b)f(ab,c) = f(b,c)f(a,bc)$ when $abc \ne 0$. This can be viewed as an associativity condition.  For suppose that we have a semigroup $\bM$ and coefficient ring $\bfk$, and that $f$ is a multiplicative 2-cocycle of $\bM$ with coefficients in the multiplicative group $\bfk^{\times}$ of units of $\bfk$.   
With $f$ we can define a new multiplication on  the semigroup algebra $\bfk\bM$ by setting  $a\ast b = f(a,b)ab$ for all $a, b \in \bM$ and then extending this bilinearly to all of $\bfk\bM$.  The cocycle condition insures that this multiplication is again associative.  We will call this a \emph{coherent deformation} of $\bfk\bM$. Since $f$ can be multiplied by any element of $\bfk^{\times}$ it actually induces a ``one parameter'' 
family of coherent deformations parameterized by $\bfk^{\times}$.  These are not at first glance deformations in the classical sense of \cite{G:DefI} (cf also \cite{GS:Monster}), but will be shown to be closely related.  When $f$ is the coboundary of a 1-cochain, say $f =\delta g$, then the mapping of $\bfk\bM$ to itself sending $a \in \bM$ to $g(a)a$ is an isomorphism of $\bfk\bM$ with the $\ast$ multiplication to $\bfk\bM$ with its original multiplication. The deformation induced by $f$ is then called \emph{trivial}.  A study of kinds of deformations related to cocycles of higher dimensions would likely lead us into the realm of Stasheff's $A_{\infty}$ algebras, cf \cite{Stasheff:Homotopy}.

If we have an additive cocycle $F$ of $\bM$ with coefficients in $\R$, then 
for all $\lambda \in \R$ we can define a multiplicative 2-cocycle $f$ by setting $f(a,b) = \exp (\lambda F(a,b))$ and obtain thereby a family of coherent deformation of $\R\bM$. We will call $f(a,b)$ a \emph{weight} that has been put on the product $ab$. More generally, suppose that we have an additive 2-cocycle $F$ of $\bM$ with coefficients not in $\R$, but in $\R /\tau\R$ for some taille $\tau$.  For every $n\in \Z$ we then have a well-defined multiplicative 2-cocycle $f$ defined by setting $f(a,b) = \exp((2n\pi i/\tau)F(a,b))$. The resulting family of coherent deformations is now quantized; the \emph{quantum number} of this $f$ is $n$.  In this case $|f(a,b)| = 1$ for all $a, b \in \bM$ and $f(a,b)$ may be interpreted as a phase. Having quantum number $n$ is equivalent to having taille equal to a submultiple $\tau/n$ of the original taille $\tau$.

To see the connection with classical algebraic deformation theory introduced in \cite{G:DefI}, observe first that a cochain $\hat f$ in the Hochschild cochain complex $C^n(\bfk\bM,\bfk\bM)$ is completely determined by its values $\hat f(a_1,\dots, a_n)$ with $a_1,\dots,a_n \in \bM$, and conversely, giving those values defines a cochain.  Starting with a semigroup cochain $f\in C^n(\bM,\bfk^{\times})$, one can define an $n$-cochain $\hat f$ in  $C^n(\bfk\bM,\bfk\bM)$ by setting $\hat f(a_1,\dots, a_n)= f(a_1,\dots, a_n)a_1a_2\cdots, a_n$.  These cochains form a subcomplex of $C^n(\bfk\bM,\bfk\bM)$ and it is easy to check that we have a cochain mapping. Conversely, those Hochschild $n$-cochains $\hat f$ such that $\hat f(a_1,\dots, a_n)$ is of the form $\lambda a_1a_2\dots a_n, \,\lambda \in \bfk^{\times}$, for all $a_1,a_2,\dots, a_n \in \bM$ form a subcomplex of the Hochschild cochain complex $C^n(\bfk\bM,\bfk\bM)$.
Sending $\hat f$ to $f \in C^n(\cM,\bfk^{\times})$ defined by setting $f(a_1,\dots,a_n) = \lambda$ is the inverse map. 

If $A$ is an arbitrary associative algebra then it was shown in \cite{G:DefI} that its second Hochschild cohomology group $H^2(A,A)$ with coefficients in itself is the set of infinitesimal deformations of $A$. It follows that the elements of $H^2(\bM,\bfk^{\times})$ can also be viewed as infinitesimal deformations of $\bfk\bM$ in the sense of \cite{G:DefI}. It is often advantageous to compute the Hochschild cohomology of an algebra not from the full Hochschild cochain complex, but by using a subcomplex defined by taking the cohomology relative to a separable subalgebra, cf. e.g. \cite{GS:SC=HC}.  Using this technique it is known that for finite poset algebras, which are in particular semigroup algebras, the inclusion of the subcomplex of the Hochschild complex just defined into the full Hochschild complex is a quasiisomorphism, i.e., induces an isomorphism in cohomology. Because of the absence of a unit and the presence of infinitely many idempotents in the path algebra, there is no guarantee that here the inclusion of this subcomplex into the full Hochschild complex induces an isomorphism of  cohomology groups. However, we conjecture that it does, when in the latter one takes suitably defined continuous cochains, noting that  $\bfk\bM$ carries a topology. If so, then $H^2(\bM, \bfk^{\times})$ would in fact be the full group of infinitesimal deformations of $\bfk\bM$ when the latter is naturally considered as a topological algebra.

If $B$ is a subalgebra of an algebra $A$ then in general a deformation of $A$ need not induce a deformation of $B$, since the expression for the deformed product of two elements in $B$ may involve elements of $A$ not contained in $B$, rather than being expressible solely in terms of elements of $B$. By contrast, suppose that  $\bM'$ is a subsemigroup of a semigroup $\bM$, that we have some coefficient ring $\bfk$, and that we have a coherent deformation of the semigroup algebra $\bfk\bM$ given by a 2-cocycle $f:\bM\times\bM \to \bfk^{\times}$. Since the values of $f$ lie in $\bfk^{\times}$ and do not involve $\bM$, it is clear that the restriction of $f$ to $\bM' \times \bM'$ defines a deformation of $\bfk \bM'$. It is also clear from the definitions that if a deformation of $\bfk \bM$ is trivial then so is the induced deformation of $\bfk \bM'$ (but not conversely). It follows that if the restriction of the deformation to $\bfk\bM'$ is not trivial then the deformation of $\bfk\bM$ itself is not trivial. This will be used later to prove the non-triviality of the deformations defined here of the path semigroup of $\cM$.

\section{Deformation of the path algebra} 

To every closed 2-form $\omega$ on $\cM$ we want to associate a 2-cochain $\tilde\omega$ of the path algebra $\cA(\cM)$.   It is only necessary to define  $\tilde\omega(\g,\g')$ for pairs of paths $\g,\g'$ in $\cM$ since the paths form a multiplicative basis for the path algebra. However, this will not be possible for every pair of paths. Suppose that $\g$ is a path from a point $a$ to $b$ and $\g'$ a path from $b'$ to $c$.  If $b \ne b'$ then set $\tilde\omega(\g,\g') = 0$. If $b=b'$, then to make the definition we will need simultaneously that there is a unique shortest geodesic $\z$ homotopic to $\g$, that there is a unique shortest geodesic $\z'$ homotopic to $\g'$ and that there is a unique shortest geodesic $\z''$ homotopic to their concatenation $\g\g'$; if these conditions are not met, then $\tilde\omega(\g,\g')$ will be left undefined. 
When the conditions are met, note that $\z''$ must also be homotopic to the concatenation of $\z$ and $\z'$. The homotopy then provides an element of area bounded by $\z, \z'$ and $\z''$. It is oriented by taking $a, b$ and $c$ in that order on the boundary, where $a, b$ are the starting and ending points, respectively, of $\g$, and $b, c$ are those of $\g'$. Then $\tilde\omega(\g,\g')$ is defined to be the integral of $\omega$ over this element. 
(More precisely, the homotopy is a mapping from the unit square into $\cM$ and one can integrate the pull back of $\omega$ over the square.) 

Note, however, that while the integral defining $\tilde\omega(\g,\g')$ is over a triangle whose sides are uniquely defined geodesics which depend only on the homotopy classes of $\g$ and $\g'$, the homotopy defining the element of area is not unique. If we have two distinct homotopies then they in effect define a mapping of the 2-sphere $S^2$ into $\cM$. The difference between the integrals will be zero if this mapping is homotopic to zero but possibly not otherwise. \emph{There is, therefore, a fundamental condition that must be imposed on $\omega$, namely, that it have a taille $\tau$. }For then the difference between the integrals will be a multiple $\tau$ and the integral becomes well-defined and independent of the choice of homotopy if reduced modulo $\tau$.  Thus $\tilde\omega$ must be understood as having values in $\R/\tau\R$.

This $\tilde\omega$ will prove to be a cocycle. The fact that $\tilde\omega(\g,\g')$ may occasionally be undefined presents no serious problem if the set of cases in which it occurs is in some sense small.   It  is natural to enlarge the concept of an algebra to allow that products be undefined in a small set of cases, and similarly for morphisms, cochains, and similar constructs. 
Since $\cM$ carries a volume form and hence a measure, so does $\cM \times \cM$. One ``smallness'' condition might be that the set of pairs of points $a,b$ such that some homotopy class of paths between them contains no unique shortest geodesic should have measure zero. We conjecture that this is in some sense almost always the case, if not always the case, but that might not be adequate. It can happen that $\cM$ has dimension 2 and that inside the 4-dimensional manifold $\cM \times \cM$ the set of such pairs of points has dimension 3, for example, the plane with a well. One would want, at least, that it not disconnect $\cM \times \cM$, but what more may be needed is unclear.

Now  suppose that we have three paths $\g, \g', \g''$ such that the products $\g \g',\g' \g''$ and $\g \g'\g''$ are all defined, and consider $\tilde\omega$ for the moment as having values (as originally) in $\R$. The coboundary of $\tilde\omega$ evaluated on these three paths is 
\begin{equation*}
\delta \tilde\omega(\g,\g',\g'')=\tilde\omega(\g',\g'') -\tilde\omega(\g\g',\g'') +\tilde\omega(\g,\g'\g'')  -\tilde\omega(\g,\g').
\end{equation*}
This is just the integral of the closed 2-form $\omega$ over the surface of a 2-simplex in $\cM$ (which happens to have geodesic edges).   It is therefore a multiple of the taille, so after reduction $\tilde\omega$  is in fact an additive 2-cocycle.  If $\tau = 0$ then introducing a parameter $\lambda$, the multiplicative 2-cocycle $\exp(\lambda\tilde\omega)$ gives a one-parameter family of coherent deformations of the path algebra $\cA(\cM)$ of $\cM$.   If  $\tau > 0$, then the multiplicative 2-cocycles $\exp((2n\pi i/\tau)\tilde\omega)$ define a discrete family of deformations indexed by $n\in \Z$; the deformations have been quantized. 

Up to algebraic isomorphism, the deformations defined here depend only on the cohomology class  in $H^2(\cM)$ of a form $\omega$, but algebraically isomorphic deformations might have significantly different properties. To compute the necessary integrals one must choose a representative form.  When $\cM$ is Riemannian it may be appropriate to choose the unique harmonic form in the class

Part of the foregoing actually involves only the \emph{geodesic algebra} of $\cM$. That is the free module generated by all directed geodesic segments on $\cM$, where the product is zero when two can not be concatenated and is otherwise the shortest geodesic homotopic to the concatenation when that geodesic is unique; otherwise it is undefined.  One sets $\tilde\omega(\g,\g')$ equal to the integral of  $\omega$ over the element of area defined by the homotopy, reduced, as before, by the taille of $\omega$. The \emph{homotopy path algebra} of $\cM$ has as underlying module the free $\R$-module generated by triples $(a,b, [\g])$, where $(a,b)$ is an ordered pair of not necessarily distinct points of $\cM$ and $[\g]$ is a homotopy class of paths $\g$ from $a$ to $b$.  The product $(a,b,[\g])(b',c,[\g'])$ is zero if $b \ne b'$, and otherwise is $(a,c,[\g\g'])$. The path algebra maps onto the homotopy path algebra by sending every path to its homotopy class. The geodesic algebra is ``essentially'' isomorphic to the  homotopy path algebra, i.e., up to the omission of a ``small'' set, since it is just the homotopy path algebra with the multiplication left undefined when there is no unique shortest geodesic in $[\g\g']$. The definition of the homotopy path algebra is, like that of the path algebra, independent of the metric on $\cM$, but the set where the essential isomorphism is undefined does depend on the metric.  

While we have explicitly shown only for dimension $n=2$ that tailor forms give rise to additive cocycles of the semigroup of paths, which can then be exponentiated to give multiplicative ones, it is clear that the same is actually true in all dimensions. In dimension 2 the multiplicative cocycles one obtains have a natural interpretation as deformations of the path algebra. As mentioned, in higher dimensions they may involve $A_{\infty}$ algebras.

\section{The weight function on a path} 

Let $\omega$ be a de Rham 2-cocycle with taille $\tau$ on $\cM$, $\tilde\omega$ be the associated additive 2-cocycle of the path semigroup $\bM$, and $\g:[0,\ell] \to \cM$ be a path of length $\ell$. The path may be self-intersecting.  Suppose that $x \in [0, \ell]$ and that there exists a unique shortest geodesic  from $\g(0)$ to $\g(x)$ homotopic to $\g([0,x])$, i.e., to that part of the path $\g$ from $\g(0)$ to $\g(x)$.  This homotopy defines an element of area over which we can integrate $\omega$. Since this homotopy need not be unique, this integral, denote it $\phi_{\g}(x)$, is only defined modulo $\tau$.  
Let $\cS$ denote the set of those points $x \in [0, \ell]$ for which the shortest geodesic from $\g(0)$ to $\g(x)$ is not unique. Then we have a function $\phi_{\g}:[0,\ell]\backslash \cS \to \R/\tau$. If $x$ is not in  $\cS$ and not a point of self-intersection then $\phi_{\g}(x)$ depends only on $\g(x)$, so we may speak of it loosely as a function on the path. If $\tau > 0$ then we define $w_{\g}(x) = \exp(2n\pi i \phi_{\g}(x)/\tau)$ to be the \emph{weight function} on the path corresponding to quantum number $n$, generalizing the special case of the weight function in the Introduction.  (For $\tau = 0$ interpret $1/\tau$ as an arbitrary parameter $\lambda$.) When $\cS$ is discrete, its points are points of possible discontinuity of the weight function. 

Every point of $a \in \cM$ has a geodesically convex neighborhood $U$, i.e., one in which any two points are joined by a unique geodesic in $U$. If $\g$ is contained in $U$ write $\phi(\g)$ for $\phi_{\g}(\ell)$. The restriction of $\omega$ to $U$ has taille 0 so this is in $\R$.  If $\g'$ is a path concatenatable with $\g$ and also contained in $U$ then we have 
$$\tilde\omega(\g,\g) = \phi(\g\g') - \phi(\g) -\phi(\g'),$$
so locally the cocycle $\tilde\omega$ is a coboundary.  Within the radius of injectivity of $\g(0)$ (the largest radius for which the exponential map at $\g(0)$ is a diffeomorphism,  equivalently, the distance to the cut locus, roughly, the set of other points to which there are multiple shortest geodesics) the  function $\phi_{\g}(x)$ is still well-defined modulo $\tau$ for  all $x\in [0,\ell]$.  However, it may no longer be a coboundary since within that radius there may be points between which there is no shortest geodesic.

Suppose that we have a deformation of the path algebra induced by a multiplicative 2-cocycle  $f$, and that a particle that has  traversed some fixed geodesic $\g$ from $a$ to $b$ is then deflected (e.g., by performing an observation or receiving an impulse) at the point $b$, continuing along a new geodesic $\g'$. Let the geodesic $\g'$ be parameterized by its length.
After traveling on $\g'$ for a distance $x$ the particle will have arrived at a point $\g'(x)$. The weight $f(\g,\g'(x))$ may be viewed as a function attached to the particle as a result of its travel.  (We have not defined the weight while the particle was on $\g$, but only once it is on the second geodesic $\g'$. However, $\g$ might be reduced to a point.)  If $f$ has been obtained from a form $\omega$ with positive taille then $|f(\g,\g'(x))| = 1$ for all $x$, so $f(\g,\g'(x))$ can be interpreted as a phase. The phase angle generally is not a linear function of $x$. It can vary discontinuously and it may happen that $f(\g,\g'(x))$ takes on only the values $\pm1$, cf \S \ref{sphere}. If this can happen with the de Broglie wave of a moving particle then points where it changes abruptly or is discontinuous may be ones where there is an observable change in the nature of the particle.

For convenience we have used length as a functional on paths but note again that there are other possibilities such as action in Lagrangian mechanics.  Also, geodesics should probably be viewed in space-time. 

Mathematically, phase space is the cotangent bundle of configuration space, and as such, its homotopy groups are those of its base, the configuration space. While these may not vanish, the elements of $\pi_2$ are all completely isotropic relative to the canonical symplectic form $\omega$, so if $\pi \in \pi_2$ then $\omega(\pi)= 0$.  Nevertheless, the existence of Planck's constant suggests that phase space must be treated as though it has a positive taille which is a submultiple of $h$. This raises the question of whether physical phase space actually does have holes.  The existence of even a single $\pi \in \pi_2$ with $\omega(\pi) \ne 0$ would force a positive taille, provided that the taille is defined, i.e., that the image of $\pi_2$ in $\R$ is discrete. Nothing precludes that $\pi_2$  even be infinitely generated, or that physical phase space consist of nothing more than a lattice of infinitesimal  bubbles scaled by $h$. In any case, the non-commutativity of its coordinates makes it behave so.

\section{Closed paths, Bohr atom}\label{Bohr}  Suppose, as before, that $\omega$ is a closed 2-form and let $\g$ be a closed path around which a particle travels periodically and on which the weight function is well-defined.  This will be the case if, for example, $\g$ is contained in a small neighborhood. Since the path is closed, the integral determining the total change of phase in traversing the path is simply that of $\omega$ over the 2-cell whose boundary is the closed path; the only function of the assumption about the weight function being well-defined (a condition on the existence of geodesics) was to insure that the path did indeed bound a 2-cell. This integral does not depend on the metric. We can therefore define it for any contractible closed path.

Suppose that $\omega$ has a positive taille $\tau$.   For a periodic orbit, a natural restriction is that  
 it return to the start with the same phase. This is precisely Bohr's hypothesis in his model of the atom.

\section{The unit sphere}\label{sphere}

Suppose, improbably, that the phase space of a particle whose configuration space was a circle, instead of being a cylinder actually closed up and became a sphere.
To show that the de Broglie wave in such a  space would not look like the usual, we discuss the deformation of the path algebra of the unit sphere.  (Consider the longitudinal direction as momentum.)
On the unit sphere $S^2$ with the usual metric, geodesics between antipodal points are not unique so the product of concatenatable paths will be undefined when the beginning of the first is antipodal to the end of the second. However, the set of pairs of antipodal points in $S^2 \times S^2$ is small; it has codimension two since the set of antipodal points is an image of the sphere inside the product of the sphere with itself. The usual area form $\omega$ is, up to constant multiple, the only harmonic $2$-form and is a natural choice of form. Its taille is the area of the sphere, namely $4\pi$.  Deformations of the path algebra here are necessarily  quantized and given by the multiplicative cocycles $\exp((n/2)i\tilde\omega), n \in \Z$.  


Suppose now that we have a deformation with quantum number $n$. Consider a particle starting in the northern hemisphere of the unit sphere at a point at 0 degrees longitude which moves south on that meridian to the equator (instantaneously acquiring momentum) and then is deflected eastward, continuing to travel on the equator. As it continues to circle the equator it will experience a change in phase, the total change in angle on returning to the equatorial point at 0 degrees longitude at which it was deflected being $n\pi$. This change is not a linear function of the distance traveled along the equator unless the particle started at the north pole. The change is most rapid as the particle passes the point on the equator at longitude 180 degrees. If the initial path to the equator had length zero, i.e., if the particle started on the equator, then the phase angle is  0 until it reaches the antipodal point, when it becomes undefined. Thereafter, until it returns to the starting point, the phase angle is $n\pi$.  If the quantum number is even, then the angle remains constant at zero modulo $2\pi$.  If the quantum number is odd, however, then there is a sign that attaches to the particle which is $+1$ from the start until the particle reaches the antipodal point, where it undefined, and then switches to $-1$ until the particle returns to the start.  At this point the particle has completed a path of length $2\pi$ but the sign does not yet return to $+1$. It is still $-1$ as the particle continues until it comes once again to the antipodal point where the sign is again undefined. Thereafter it remains  $+1$ as it passes the starting point, switching every time it passes the antipodal point.  It must travel twice around the equator until it returns to its original state. The sign attached to a particle simply circling the equator switches exactly once in a full orbit, independent of the value of $n$, as long as $n$ is odd. The changes in sign occur even though the particle has not been deflected. By contrast, if the particle has not started at the equator but has been deflected there to travel along the equator, then the change in phase depends on $n$.  For even values of $n$, particles which are not deflected experience no change in phase, so the deformation is not apparent until they are deflected.  In either case, particles starting at a common point and ending at a common point may arrive with different phases, depending on their paths.

\section{Non-triviality of deformations} The possible physical implications of deformation of the path algebra of phase space may not depend on the non-triviality of the deformation.  When a closed but not exact tailor form $\omega$ has taille equal to zero we will see that the one-parameter family of deformations which it defines  is non-trivial because its infinitesimal is non-trivial. However, if 
$\pi_2 \ne 0$ and the taille is consequently positive, then the deformations induced by $\omega$ are quantized and one has, in effect, only certain discrete specializations of what would have been a continuous family. These may conceivably be trivial even when the family is not, cf \cite{GerstGiaq:Rigid}. However, the deformations constructed here are non-trivial when the topology is considered. 

It is a classic theorem that a differentiable manifold $\cX$ has a triangulation, that is, that there is a simplicial complex $K$ homeomorphic to  $\cX$ together with a homeomorphism $\theta:K \to \cX$.  This applies, in particular to $\cM$. The theorem is due  J. H. C. Whitehead \cite{JHCWhite:Triang} but based on earlier work of S.S. Cairns; for a brief history of the ideas cf the thesis of Emil Saucon \cite{Saucon:Munkres}. A manifold which is not differentiable need not have a triangulation.

Consider now a deformation of the path algebra of $\cM$ induced by  some non-trivial de Rham 2-cocycle $\omega$. Let $\tilde\omega$ denote the additive 2-cocycle  which $\omega$ defines on the cochain complex of the semigroup $\bM$ of the paths of $\cM$.  If the taille $\tau$ of $\omega$ is zero then we actually have a one-parameter family of deformations given by $\exp(\lambda \tilde\omega)$, where $\lambda$ is the parameter. To show that the family whose infinitesimal is $\tilde\omega$  is non-trivial it is then sufficient to show it is so for the restriction to some subsemigroup.
Let $K$ be a triangulation of $\cM$.  Then there is a 2-cycle $\zeta$ of $K$ such that the integral of $\omega$ over the image $\theta(\zeta)$ of $\zeta$ is not zero. The barycentric subdivision $\zeta'$ of $\zeta$ is a partially ordered set. The path semigroup generated by the images of its vertices and its 1-cells is a finite subsemigroup $\bM'$ of the path semigroup $\bM$ of $\cM$. We can restrict $\tilde\omega$ to this subsemigroup, where it remains non-trivial because the integral of $\omega$ over $\theta(\zeta')$, which is the same as its integral over $\theta(\zeta)$, not zero.  In this case the deformation of the path algebra of $\cM$, considered as an abstract algebra without regard to its topology, is non-trivial.

If the taille of $\omega$ is $\tau > 0$ then the deformation induced by the additive 2-cocycle $\tilde\omega$ is given by a multiplicative 2-cocycle of the form $f = \exp((2n\pi i/\tau) \tilde\omega)$ for some integer $n$ which we may assume is not zero.  While $\tilde\omega$ is non-trivial as an additive cocycle of $\bM$ it may be that when viewed as having coefficients in $\R/\tau$ it becomes a coboundary of the path semigroup $\bM$ when the latter is considered without its topology. (This is the case in the example of the 2-sphere.)  That is, we may be able to find a 1-cochain $\phi$ such that $\tilde\omega(\g,\g') = \phi(\g)+\phi(\g')- \phi(\g\g') \mod\tau$ for all paths $\g, \g' \in \cM$.  Note that $\tilde\omega$, having been defined as an integral, is automatically continuous. We will show that no such continuous $\phi$ can exist. Choosing an arbitrary $\epsilon > 0$ we may, using repeated barycentric subdivision, assume that we have taken a triangulation so fine that the length of no path in $\bM'$ (which is finite) is  greater than  $\epsilon$.  Since $\phi$ is continuous, we may choose $\epsilon$ so small that $|\phi(\g)| < \epsilon/3$ for all paths $a$ on $\bM'$. The preceding equation can then hold only if  $\tilde\omega(\g,\g') = \phi(\g)+\phi(\g')- \phi(\g\g')$ for all $\g,\g'$, implying that $\tilde\omega$ was a coboundary in $\bM'$.   However, the cohomology of $\bM'$ is still that of the image of the original $\zeta$, no matter how fine the subdivision, a contradiction.


\end{document}